\documentclass[a4paper]{article}
\usepackage{INTERSPEECH2018}
\usepackage{mathtools}
\usepackage{graphics}
\usepackage{dirtree,balance}

\title{The VOiCES from a Distance Challenge 2019 Evaluation Plan}
\name{Mahesh Kumar Nandwana$^1$, Julien Van Hout$^1$, Mitchell McLaren$^1$, Colleen Richey$^1$, \\ Aaron Lawson$^1$, Maria Alejandra Barrios$^2$}
\address{
  $^1$Speech Technology and Research Laboratory, SRI International, Menlo Park, California, USA\\
  $^2$Lab41, In-Q-Tel Laboratories, Menlo Park, California, USA  \newline Updated: \bf\today}
\email{voices\_poc@sri.com}

\begin{document}
\maketitle

\section{Introduction}
The ``VOiCES from a Distance Challenge 2019'' is designed to foster research in the area of speaker recognition and automatic speech recognition (ASR) with the special focus on single channel distant/far-field audio, under noisy conditions. The main objectives of this challenge are to: (i) benchmark state-of-the-art technology in the area of speaker recognition and automatic speech recognition (ASR), (ii) support the development of new ideas and technologies in speaker recognition and ASR, (iii) support new research groups entering the field of distant/far-field speech processing, and (iv) provide a new, publicly available dataset to the community that exhibits realistic distance characteristics.

This challenge is based on the recently released Voices Obscured in Complex Environmental Settings (VOiCES) corpus, released under Creative Commons-BY 4.0 license, making it accessible for commercial, academic, and government use. The VOiCES corpus, a collaboration between SRI International and Lab41, provides speech data recorded in acoustically challenging environments. Data was collected by recording retransmited audio from high-quality loudspeakers that played in real rooms, capturing natural reverberation. LibriSpeech~\cite{LibriSpeech2015} was used as the clean speech source, while television, music or babble played simultaneously from another loudspeaker as background noise. The clean speech loudspeaker rotated at predefined intervals during recordings, to mimicking human head movement. More details on VOiCES can be found at~\cite{voices,nandwana2018robust}.

Participation in the VOiCES challenge is free of cost. The challenge is intended for those interested in upholding the challenge rules outlined in this document and who intend to submit a paper to ``The VOiCES from a Distance Challenge 2019'', a Special Session to be held at Interspeech 2019, in Graz, Austria on September 15-19, 2019. Information about evaluation registration can be found on the VOiCES website\footnote{https://voices18.github.io/Interspeech2019\_SpecialSession/}.

Participants who complete the challenge and submit both their system outputs and description will get early access to the VOiCES phase 2 data. The phase 2 data is an extension of VOiCES phase 1 set,  having over 310k audio files recorded in different reverberant environments.

The VOiCES challenge has two tasks: speaker recognition and automatic speech recognition (ASR). Each task has fixed and open training conditions. These conditions are defined by the training data that can be used to train the systems. Participants are required to participate in at least one condition of a task (e.g. ASR Open).

\section{Speaker Recognition}

The speaker recognition challenge presented here is similar to previous speaker detection challenges, such as the National Institute in Standards of Technology (NIST) Speaker Recognition Evaluations (SRE)~\cite{Sadjadi2017} and the Speakers in the Wild (SITW) challenge~\cite{sitw}. The task is: given a segment of speech and target speaker enrollment data, automatically determine whether the target speaker is speaking in the segment. A segment of speech (test segment) and the enrollment speech segment(s) from a designated target speaker constitute a trial. The speaker recognition system is required to process each trial independently and output a log-likelihood ratio (LLR), using natural (base $e$) logarithm, for that trial. The LLR for a given trial including a test segment $s$ is defined as follows:

\begin{equation}
	LLR(s) = \log\left(\frac{P(s|H_{0})}{P(s|H_{1})}\right)
\end{equation}

where $P(.)$ denotes the probability distribution function (pdf), and $H_{0}$ and $H_{1}$ represents the null (i.e., $s$ is spoken by the enrollment speaker) and alternative (i.e., $s$ is not spoken by the enrollment speaker) hypotheses, respectively. The performance of a speaker recognition system will be judged on the accuracy of these LLRs.

\subsection{Training Conditions}
Speaker recognition systems can be developed for the fixed condition, the open condition or both. The two training conditions are defined by the specific datasets that can be used to build the speaker recognition system. 

\subsubsection{Fixed Condition}
The fixed training condition limits the system training to the following freely available data sets:

\begin{itemize}
	\item Speakers in the Wild (SITW)\footnote{http://www.speech.sri.com/projects/sitw/}
	\item VoxCeleb1~\cite{Nagrani2017} and VoxCeleb2\footnote{http://www.robots.ox.ac.uk/~vgg/data/voxceleb/}~\cite{Chung2018}
\end{itemize}
 
 Participants can obtain these datasets by following the instructions on their respective webpages. The audio data from the Voxceleb1 and Voxceleb2 is restricted to the official annotations for the fixed condition submissions. In this way, the fixed condition can serve its purpose of measuring the performance of different systems trained with the same data (or a subset thereof). The Voxceleb datasets also contain video URLs. No image or video processing is allowed in the fixed condition, however, image or video processing may be used for cross-model processing of the audio used to train a system for the open condition. Please note that SITW and VoxCeleb have overlapping speakers\footnote{http://www.robots.ox.ac.uk/~vgg/data/voxceleb/SITW\_overlap.txt}.
 
 Publicly available non-speech audio and noises (e.g. noises, impulse responses, codecs) may be used for data augmentation~\cite{x-vectors,McLaren2018} and should be included in the system description. For the fixed training condition, only the datasets specified above maybe used for system training and development with the exception of speech activity detection (SAD). Participants may train their own or use an existing SAD and details of the SAD should be included in the system description. With the exception of SAD, participants in the fixed condition can not use a pre-trained model for system components.

\subsubsection{Open Condition}
The open training condition removes the limitations of the fixed condition. For this condition, participants can use any proprietary and/or public data they have access to including the fixed condition data. The participants must mention the datasets used to train the open condition submission in the system description.

\subsection{Development Data}
The speaker recognition development dataset consists of 15,904 audio segments from 196 speakers. Each audio file contains a single speaker. The dataset represents different rooms, microphones, noise distractors, and loudspeaker angles. The metadata is available in the filename and teams may use this information to analyze the behavior of their system under different conditions. More detailed information about the metadata can be found in the README.VOiCES\_2019.txt provided with the development data.

The development data for speaker recognition may be used for system training including the calibration models for both fixed and open conditions. 

\subsection{Evaluation Data}

The speaker recognition evaluation set will consist of unreleased distant recordings that are part of the VOiCES corpus. Participants can expect these recordings to originate from different microphone types and different rooms both of which could be more challenging than those featured in the development set.

\subsection{Performance Measures}
We will use several performance measures to determine the speaker recognition system performance and compare system submissions in the challenge. A metric similar to those used in the NIST SRE 2010 and Speakers in the Wild (SITW)~\cite{sitw} challenge will form the primary metric for the VOiCES challenge. 

The participants have been provided with a python script to evaluate each performance metrics detailed below. This python scorer will be used by the organizers to produce the official metrics on the evaluation data. 

\subsubsection{Primary Metric}
The primary metric for the VOiCES challenge is based on the following detection cost function, which is the same function used in the NIST 2010 SRE, but with different parameters. It is a weighted sum of miss and false alarm error probabilities in the form:

\begin{equation}
 C_{det} = C_{miss} \times  P_{miss} \times P_{tar} + C_{fa} \times P_{fa} \times (1-P_{tar})
\end{equation}

We assume a prior target probability, $P_{tar}$, of 0.01 and equal costs between misses and false alarms. 

\begin{table}[h]
	\centering
	\caption{Cost model parameters for the primary metric $C_{det}$}
	\begin{tabular} {clclc}
		\toprule
		$C_{miss}$& $C_{fa}$& $P_{tar}$ \\
		\midrule
		1.0 & 1.0 & 0.01 \\
		\midrule
	\end{tabular}
\end{table}

For reporting, the $C_{det}$ will be normalized by the cost that a na\"ive system that always chooses the least costly class would get for the selected parameters. In our case, the normalization factor is given by $P_{tar}$.

\subsubsection{Alternate Performance Metric}

For the purpose of analyzing how well a system is calibrated across all operating points, a log-likelihood ratio cost metric, $C_{llr}$~\cite{niko_cllr}, will also be reported as:
\begin{equation}
\scalebox{0.95}[1]{$C_{llr} = \frac{1}{2 \times \log(2)} \times \left( \frac{\sum \log(1+1/s)}{N_{tar}} + \frac{\sum \log(1+s)}{N_{non}}\right)$}
\end{equation}
where $s$ is the likelihood ratio for a trial, and $N_{tar}$ and $N_{non}$ represent the number of target and non-target trials, respectively. 

\subsection{Scores Submission}

Participants are required to submit to the VOiCES organizers a set of scores for each trial they evaluated. The score files should follow the naming convention: [TeamName]\_[Task]\_[Condition]\_[SystemNumber].txt.

The score files should be in ASCII format with one line per trial. Each line must include three space-delimited fields:
modelID$<$space$>$testSegment$<$space$>$LLR$<$NewLine$>$

A separate score file is required for each condition and each system, with a limit of three files per condition. The score submission instructions will be provided along with the evaluation data.

\section{Automatic Speech Recognition}

In the ASR task, participants are expected to provide a transcript of each audio segment in a verbatim and case-insensitive manner. 

\subsection{Training Conditions}

The ASR task will be evaluated over fixed and open training conditions. The two training conditions are defined by the specific datasets that can be used to build the speech recognition system. Teams can participate in the fixed condition, open condition or both.

\subsubsection{Fixed Condition}

In the fixed condition, the training set consists of an 80-hour subset of the LibriSpeech corpus. This subset was designed in such a way as to have no overlap in speakers with the VOiCES corpus (dev or eval).

While the participants may train their own SAD as well as use external non-speech resources for data augmentation, they may not use additional speech data from any other source for model training (acoustic model, language model, speech enhancement, etc.)

\subsubsection{Open Condition}

The open condition removes the limitations of the fixed condition. For this condition, participants can use any proprietary and/or public data they have access to along with the fixed condition data.

\subsection{Development Set}

The ASR development set is distinct from the speaker recognition dev set, and will consist of 20h of distant recordings from rooms 1 and 2 along with corresponding transcripts. It contains recordings from 6 of the 12 mics, and is balanced across rooms, mics, distractor types, and loudspeaker angles. The metadata (mic, room, distractor, angle) is available in the filename, and the participants are welcome to use that information to analyze the behavior of their system under different conditions.

{\bf The VOiCES Challenge's dev set may be used to make design decisions, but may not be used for directly training the system's SAD, enhancement, acoustic or language models in either the fixed or the open condition.}

\subsection{Evaluation Set}

The ASR evaluation set will consist of 10h to 20h of previously unreleased distant recordings that are part of the VOiCES corpus. Participants can expect the recordings to originate from different microphone types and different rooms, both of which could be more challenging than those featured in the dev set.

\subsection{Performance Measures}

Similarly to NIST's recent OPENSAT-17 challenge, we will use the Word-Error Rate (WER) as the evaluation metric for the ASR portion of this challenge. Specifically, we will use NIST's open source SCTK software to score participants submissions by computing WER as the sum of errors (deletions, insertions, substitutions) divided by the total number of words from the reference transcript.

\subsection{Scores Submission}

Participants are required to submit to the VOiCES organizers word-level transcripts in Conversation-Time Marked (CTM) format. The score files should follow the naming convention: [TeamName]\_[Task]\_[Condition]\_[SystemNumber].txt.

The CTM format consists of a tab-separated 6-columns ASCII text file, where each line corresponds to a word. The fields are defined as follows:
\begin{enumerate}
	\item The waveform file base name (i.e., without path names or extensions).
	\item Channel ID, the audio files are mono this column should be `1'
	\item The beginning time of the word, in seconds, measured from the start time of the file.
	\item The duration of the word, in seconds
	\item The orthographic rendering (spelling) of the token.
	\item Confidence Score, the probability with a range [0:1] that the token is correct. If confidence is not available, omit the column.
\end{enumerate}

The ASR scoring is identical to the NIST OPENSAT-17 evaluation, and more details can be found in NIST's evaluation plan\footnote{https://www.nist.gov/itl/iad/mig/opensat}.

A separate score file is required for each condition and each system, with a limit of three files per condition. The score submission instructions will be provided along with the evaluation data.

\section{Training and Evaluation Dataset Organization}

The data structure of both speaker recognition and speech recognition data download is as follows: 

\dirtree{%
	.1 Interspeech2019\_VOiCES\_Challenge/. 
	.2 Training\_Data.
	.3 Automatic\_Speech\_Recognition.
	.2 Development\_Data.
	.3 Speaker\_Recognition.
	.3 Automatic\_Speech\_Recognition.
	.2 Evaluation\_Data.
	.2 README.VOiCES\_2019.txt.
}

The meta information of the audio files is included in the file names. More information about metadata can be found in README.VOiCES\_2019.txt

\section{Evaluation Rules}

All participants must adhere to the following rules regarding the processing of the VOiCES evaluation data until all system outputs have been submitted.

\begin{itemize}
	\item {\bf Participants may only use the subset of the VOiCES data provided for development and evaluation under each task. Teams may \emph{not} use any other VOiCES data releases including parts of the VOiCES corpus that contain distractor audio only. This is a challenge and using such data will give unfair advantage.}
	\item Participants may not use ASR development data for speaker recognition tasks and vice versa.
	\item Participants must submit system output for at least one task condition (i.e. ASR Open).
	\item Participants must abide by the terms guiding the fixed and open training conditions. 
	\item Participants may not probe the evaluation data via manual/human means such as listening to the data or producing the transcript of the speech. 
	\item Participants may submit up to three system per task (SID/ASR) and condition (Fixed/Open). 
	\item The official score for a team will be selected as the best primary metric from systems submitted by the team for that condition (up to 3 systems can be submitted per condition per task per team). These official scores will be used for ranking teams.
	\item Each team must submit an article describing their systems and providing analysis of its performance on the VOiCES database to the special session of Interspeech 2019 paper submission deadline. 
	\item During the challenge, teams may email questions seeking clarification of any aspects, specifically those that might be considered vague or ambiguous. To ensure all teams receive the same information, a summary of each question and the response will be emailed to the contact person of each team with the poser of the question being made anonymous.
	\item The organizers plan to write articles comparing techniques by anonymizing submissions.  Official rankings of teams will be published  on  the  VOiCES  website,  including  scores  and  confidence margins; individual team requests for anonymity on this public website will be upheld. Regarding further dissemination of results, participants are allowed to publish their own results and their rank from challenge results. They are not allowed, however, to publish other teams results or rank from the challenge. The only exception to this is when referencing published results with the corresponding team authoring such publications. 
	\item {\bf Participants are not allowed to use any part of LibriSpeech for fixed or open condition system training except for the data provided as ASR training set for the fixed condition. LibriSpeech is the data source for VOiCES data and there is a possibility that there will be an overlap between training and evaluation set.}
\end{itemize}

\section{System Description}

All participants will be required to submit a short system description by March 15, 2019. The purpose of this description is two-fold: it will be shared among other participants for the benefit of analysis and validation, and it will provide information necessary for the organizers to determine common trends in leading systems.  The organizers may then use this information (without site names for anonymity) in a ‘summary’ article submitted to the special session. System description is required to get access to the evaluation keys and VOiCES phase 2 download link.

\section{Special Session Paper at Interspeech}

In addition to the system description above, participants must submit an article to "The VOiCES from a Distance Challenge" special session track of Interspeech 2019. The information in the system description will also form part of the paper along with any post-evaluation analysis. This should adhere to the Interspeech 2019 paper submission guidelines~\footnote{https://interspeech2019.org/authors/author\_resources/} and schedule~\footnote{https://interspeech2019.org/calls/important\_dates/}.  

\section{Schedule}
Limited time is available for development due to the time between special session approval and the regular paper submission deadline. The schedule below aims to provide as much time as possible for development while also allowing sufficient time for the post-evaluation analysis for the special session and time for writing a paper for Interspeech special session.

\begin{itemize}
	\item {\bf January 15, 2019:} Release of the evaluation plan and development sets
	\item {\bf February 25, 2019:} Evaluation data available 
	\item {\bf March 4, 2019:} System output submission deadline (11:59 PM PST)
	\item {\bf March 11, 2019:} Release of the evaluation results 
	\item {\bf March 15, 2019:} System description submission and release of VOiCES phase 2 key for the participating teams 
	\item {\bf March 29, 2019:} Regular paper submission deadline for Interspeech 2019
\end{itemize}

\balance
\bibliographystyle{IEEEtran}
\bibliography{mybib}

\end{document}